# Formula-Guided Machine Learning for Ground Vibration Propagation and Attenuation Modeling

**Pei-Yao Chen[a,1], Chen Wang[a,1], Fang Yan[b,c,*], Chao-Yang Zhang[b,d], Xiang-Yu Tan[b,c], Guo-Ping Lin[b], Jian-Sheng Fan[a]**

[a] Department of Civil Engineering, Tsinghua University, Beijing 100084, China

[b] Institute of High Energy Physics, Chinese Academy of Sciences， Beijing, 100049, China

[c] University of the Chinese Academy of Sciences, Beijing, 100049, China

[d] College of Water Resources and Civil Engineering, China Agricultural University, Beijing, 100083, China

[*] Corresponding Author

E-mail addresses: pyaoch@163.com (P. Chen), chwang@tsinghua.edu.cn (C. Wang), yanfang@ihep.ac.cn (F. Yan), zhangchy@ihep.ac.cn (C. Zhang), xytan@ihep.ac.cn (X. Tan), lingp@ihep.ac.cn (G. Lin), fanjsh@tsinghua.edu.cn (J. Fan)

[1] The two authors contribute equally to this work

## Abstract

Understanding the propagation and attenuation patterns of ground vibrations is critical for evaluating the impact of environmental disturbances on large-scale scientific facilities. However, complex site conditions often result in intricate vibration behaviors, limiting the accuracy of traditional predictive methods. This study proposes a hybrid iterative fitting method that integrates machine learning with the Bornitz formula through an intelligent formula generation model. The method enables the automatic derivation of high-precision, interpretable ground vibration attenuation formulas from experimental data. A case study was conducted at the High Energy Photon Source in Beijing, where field tests were performed to collect vibration data. Using the proposed approach, an attenuation formula describing ground vibration propagation was derived. The physical validity of the model was further verified via finite element simulations. A probabilistic analysis was then employed to estimate computational errors. Comparative evaluations with black-box machine learning models and empirical formulas from previous studies demonstrate that the proposed method offers significant advantages in both interpretability and accuracy. These findings provide a valuable framework for vibration impact assessment and mitigation in other large-scale scientific infrastructure projects.

# 1. Introduction

The propagation and attenuation of vibrations have important applications in fields such as transportation [1,2], pipelines [3,4] and environmental engineering [5–7]. With the advancement of technology, the infrastructure supporting high-precision technologies demands stringent control over vibrations. Ground vibrations, a common form of vibration resulting from human activities such as vehicles and railways [8–10], have a significant impact on the performance of sensitive equipment. Therefore, understanding the propagation and attenuation mechanisms of these vibrations is essential for evaluating their effects on precision instruments. However, the complexity of soil properties and environmental factors such as reflection and interference, caused by site conditions, make the propagation and attenuation behaviors of ground vibrations highly intricate. Thus, it is necessary to conduct field experiments and derive the propagation and attenuation patterns of ground vibrations from the experimental data.

The study of ground vibration propagation and attenuation is generally approached through a combination of theoretical models and field experiments. Yang et al. [11] proposed an analytical solution for the ground vibration of a half-space induced by sub- and super-critical harmonic moving loads, employing a contour integral combined with the residue theorem. Wu et al. [12] investigated the propagation and attenuation of elastic waves in multi-row infinitely periodic pile barriers, validating their findings through several numerical cases. Wang et al. [13] characterized the vibration effects and developed a real-time statistical method for attenuation of vibration effects during continuous ground-penetrating radar surveys. The processing time of this method is within seconds, enabling real-time and automatic attenuation of vibration effects. Zarei et al. [14] comprehensively evaluates the vibrations induced by heavy vehicles on asphalt pavements, developing a 2.5D finite element model to analyze the impact of vehicle speed and bump height on ground-borne vibrations and providing a predictive model for vibration levels as a guideline for speed bump installation near sensitive structures. However, due to the complex nature of soil, which is difficult to approximate using ideal theoretical models or solve through numerical methods such as finite element analysis, data-driven end-to-end models present a more effective approach.

In recent years, the development of machine learning and artificial intelligence has introduced a new paradigm for addressing data-driven problems in the engineering field [15–17]. Many researchers have applied these technologies to the modeling and fitting of ground vibration propagation and attenuation patterns. Cao et al. [18] proposed a novel distance estimation algorithm for periodic surface vibrations, introducing a frequency band energy percentage feature and an enhanced computationally efficient k nearest neighborhood algorithm, demonstrating superior performance in distance prediction for earth surface periodic vibration signals. Monjezi et al. [19] proposed a prediction model for blast-induced ground vibration using artificial neural networks, demonstrating its superior accuracy compared to traditional empirical and statistical methods in predicting peak particle velocity based on parameters such as maximum charge per delay, distance from the blasting face, stemming, and hole depth. Liu et al. [20] proposed an efficient deep learning-based approach for identifying and analyzing train-induced ground-borne vibrations, utilizing a hybrid CNN-LSTM model and continuous wavelet transform for feature extraction, and investigates the uncertainty in vibration responses through long-term monitoring and statistical analysis. Verma & Singh [21] proposed a comparative study on the application of genetic algorithm optimization technique for predicting peak particle velocity in ground vibration, demonstrating its superior accuracy and robustness over traditional empirical equations and artificial neural networks in predicting ground vibration parameters based on blast design and explosive parameters. Although deep neural networks and other black-box models possess strong data fitting capabilities, their lack of physical interpretability raises concerns about their out-of-distribution generalization ability, particularly for problems with a strong physical basis, such as vibration propagation. This limitation makes it challenging to integrate these models with subsequent assessment tasks.

To address the low accuracy of theoretical formulas and the lack of interpretability in black-box models, this study proposes a hybrid iterative method that integrates the Bornitz theoretical formula with an intelligent formula generation model. The proposed approach preserves the interpretability of classical theoretical models while leveraging the powerful fitting capabilities of deep learning techniques. The effectiveness of this method is validated through a case study at the High Energy Photon Source (HEPS), a major national scientific infrastructure located in Huairou Science City, China. The main contributions of this work are as follows:

1. Combining traditional machine learning methods with intelligent formula generation model to discover propagation and attenuation laws based on the Bornitz model, leading to the development of a ground vibration propagation formula that is frequency-dependent.

2. Field experiments were conducted at HEPS using a vibrator to generate sinusoidal excitation forces in the frequency range of 1–100 Hz. Using the proposed method, an attenuation formula with high accuracy and strong interpretability was derived from the experimental data.

3. Providing an explanation of the physical rationality of the formula based on theoretical derivation and numerical simulation.

4. Employing probabilistic analysis methods to estimate the error range of the fitting formula and

comparing the proposed method with black-box machine learning models and previous empirical formulas, demonstrating its significant advantages in terms of accuracy and interpretability.

The organization of the remaining sections of the paper is as follows. Section 2 introduces the hybrid iterative fitting method for attenuation modeling, which combines the Bornitz formula with an intelligent formula generation model. Section 3 details the field experiments, data processing procedures, and the process of formula discovery, followed by an analysis of the physical plausibility of the results. In Section 4, a probabilistic approach is used to evaluate the error of the derived formula, and its performance is compared with that of black-box AI models and existing empirical formulas. Section 5 concludes the paper with a summary of the findings.

# 2. Formula-Guided Machine Learning Method

## 2.1. Bornitz Model

The Bornitz formula [22] is a classical model used to describe ground vibration propagation, which approximates the nonlinear behavior of soil by introducing geometric and material damping coefficients. However, the original form of the Bornitz formula is overly simplistic and does not accommodate the diverse needs of different site types. As a result, many researchers have calibrated and modified the Bornitz formula based on field test results. Kim & Lee [23] represented various types of vibration sources and their induced waves through geometric modeling, determining the geometric damping coefficients for different vibrations, which showed good agreement with field test results. Ren et al. [24] developed a vibration attenuation model in layered homogeneous media based on the Bornitz formula, providing an alternative tool to further investigate the impact of high-speed train-induced vibrations on foundation soils and adjacent underground structures. Niu et al. [25], building upon the Bornitz model and incorporating measured data, discussed the geometric attenuation and damping attenuation coefficients of the formula, using the optimized Bornitz model to predict ground vibrations caused by high-speed trains in loess areas. However, due to the limitations in the frequency range of field tests and the challenges in deriving the formula, research on frequency-dependent Bornitz models remains insufficient, failing to capture the elastoplastic characteristics of soil, such as resonance and absorption.

The propagation and attenuation of ground vibrations are influenced by various factors, including the amplitude of the vibration source, propagation distance, and soil properties. For a given vibration, the energy attenuation during propagation—manifested as amplitude decay—is primarily governed by two key factors: geometric and material attenuation. This process is well described by the Bornitz equation [22]:

$$A_r = A_0 \left( \frac{r_0}{r} \right)^n e^{-\alpha(r-r_0)} \tag{1}$$

where $A_0$ and $r_0$ represent the amplitude and distance from the vibration source at the reference point, while $A_1$ and $r_1$ denote the amplitude and distance from the source at the calculation point. The parameters $n$ and $\alpha$ correspond to the geometric damping coefficient and material damping coefficient, respectively.

## 2.2. Intelligent formula generation model

In recent years, with the advancement of generative artificial intelligence, research on its application to formula discovery tasks has been steadily emerging, leading to the development of intelligent models for formula generation [26–28]. These formula generation models represent mathematical expressions in the form of prefix notation. For example, the expression $a \times (b + c)$ is rewritten as $\times\ a + b\ c$, thereby eliminating the need for parentheses. This transformation converts a formula into a sequence composed solely of "tokens", including operators and variables, which can then be processed by generative artificial intelligence models such as Long Short-Term Memory (LSTM) networks [29] and Transformer architectures [30]. Chen et al. [31] integrated feature engineering with intelligent formula generation models, significantly enhancing the model's representational capacity. This advancement enables the generation of physically meaningful, dimensionally consistent piecewise function formulas and facilitates the handling of complex engineering problems.

However, intelligent formula generation models face computational efficiency challenges when dealing with large-scale data. Moreover, they are not well-suited for formula discovery tasks of where part of the equation is fixed while the other part requires fitting. To address these limitations, we build upon the model proposed by Chen et al. [31] and introduce a hybrid iterative fitting approach that integrates traditional machine learning with intelligent formula generation models, specifically designed for the formula discovery task based on Bornitz model.

## 2.3. Hybrid iterative fitting approach

As a fundamental equation describing the attenuation law, the Bornitz equation is independent of the choice of the reference point, ensuring that the following three relationships remain consistently valid for any $r_0$, $r_1$ and $r_2$:

$$A_{r_1} = A_0 \left( \frac{r_0}{r_1} \right)^n e^{-\alpha(r_1 - r_0)} \tag{2}$$

$$A_{r_2} = A_0 \left( \frac{r_0}{r_2} \right)^n e^{-\alpha(r_2 - r_0)} \tag{3}$$

$$A_{r_2} = A_{r_1} \left( \frac{r_1}{r_2} \right)^n e^{-\alpha(r_2 - r_1)} \tag{4}$$

This self-consistency arises from the inherent properties of power functions and exponential functions, which also reinforce the physical significance of the Bornitz equation. To account for the influence of frequency on vibration propagation and attenuation while maintaining this self-consistency, we treat the geometric damping coefficient $n$ and the material damping coefficient $\alpha$ as functions of frequency $f$. Consequently, we derive the functional form to be fitted as follows:

$$A_r = A_0\left(\frac{r_0}{r}\right)^{n(f)} e^{-\alpha(f)(r-r_0)} \tag{5}$$

For a given frequency $f$, a simple transformation of Equation (5) yields the following relationship:

$$n(f)\log\left(\frac{r_0}{r}\right) = \log(\frac{A_r}{A_0}) + \alpha(f)(r - r_0) \tag{6}$$

$$\alpha(f)(r - r_0) = n(f)\log\left(\frac{r_0}{r}\right) - \log(\frac{A_r}{A_0}) \tag{7}$$

In other words, if $\alpha$ (or $n$) is treated as a constant, determining $n$ (or $\alpha$) at a specific frequency $f$ can be formulated as a linear fitting problem without bias. This type of problem can be described as follows: Given a set of data points $\{(x_1, y_1), (x_2, y_2), \ldots, (x_n, y_n)\}$, the goal is to find the optimal $k$ such that the fitting relationship $kx = y$ minimizes the squared error. Through a straightforward derivation, it can be obtained that:

$$k = \frac{\sum_{i=1}^{n} x_i y_i}{\sum_{i=1}^{n} x_i^2} \tag{8}$$

Thus, without altering the values of $\alpha(f)$ (or $n(f)$), we can determine the optimal $n$ (or $\alpha$) for each frequency $f$ that minimizes the fitting error. Next, we employ the intelligent formula generation model to identify the functional relationship $n(f)$ (or $\alpha(f)$). We then reverse the process, fixing the obtained $n(f)$ (or $\alpha(f)$) and repeating the same procedure to derive a new functional relationship for $\alpha(f)$ (or $n(f)$). Through iterative refinement, the formula's fitting error gradually decreases, ultimately yielding a highly accurate fitted expression. Starting with $n(f) = 0$ as the initial condition, the workflow of our method is illustrated in **Fig. 1**.

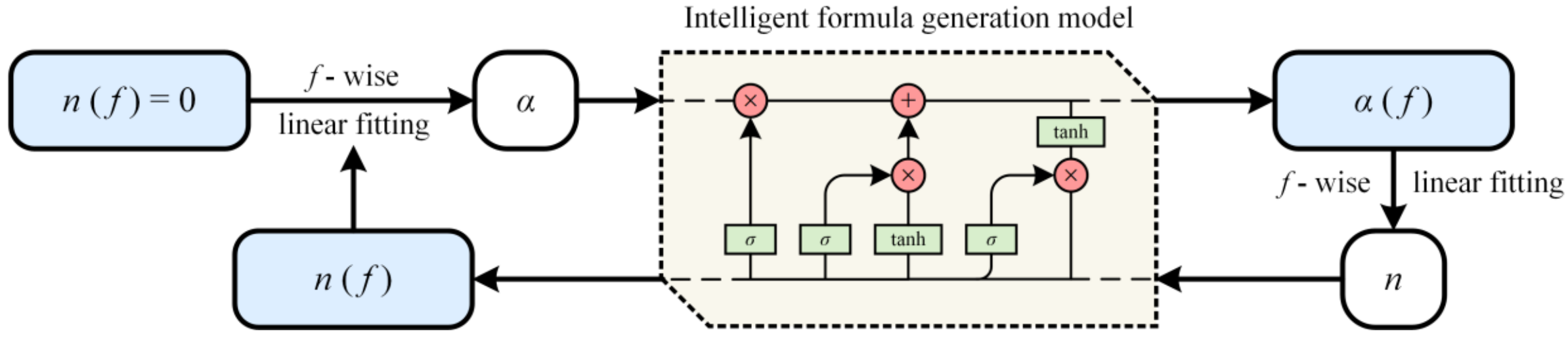

**Fig. 1** The workflow of our formula discovery method

## 3. Experiment and results

### 3.1. Field test on HEPS

The High Energy Photon Source (HEPS), as shown in **Fig. 2**, is the first fourth-generation synchrotron light source facility being built in China [32]. Located in the northern core area of Huairou Science City in Beijing, this advanced facility serves as a high-performance, high-energy synchrotron radiation light source. Its high luminosity design demands stringent beam stability, which in turn requires precise control over surrounding vibrations. Consequently, it is crucial to investigate the vibration attenuation and propagation characteristics at the HEPS site to evaluate the impact of ground vibrations on the facility's performance.

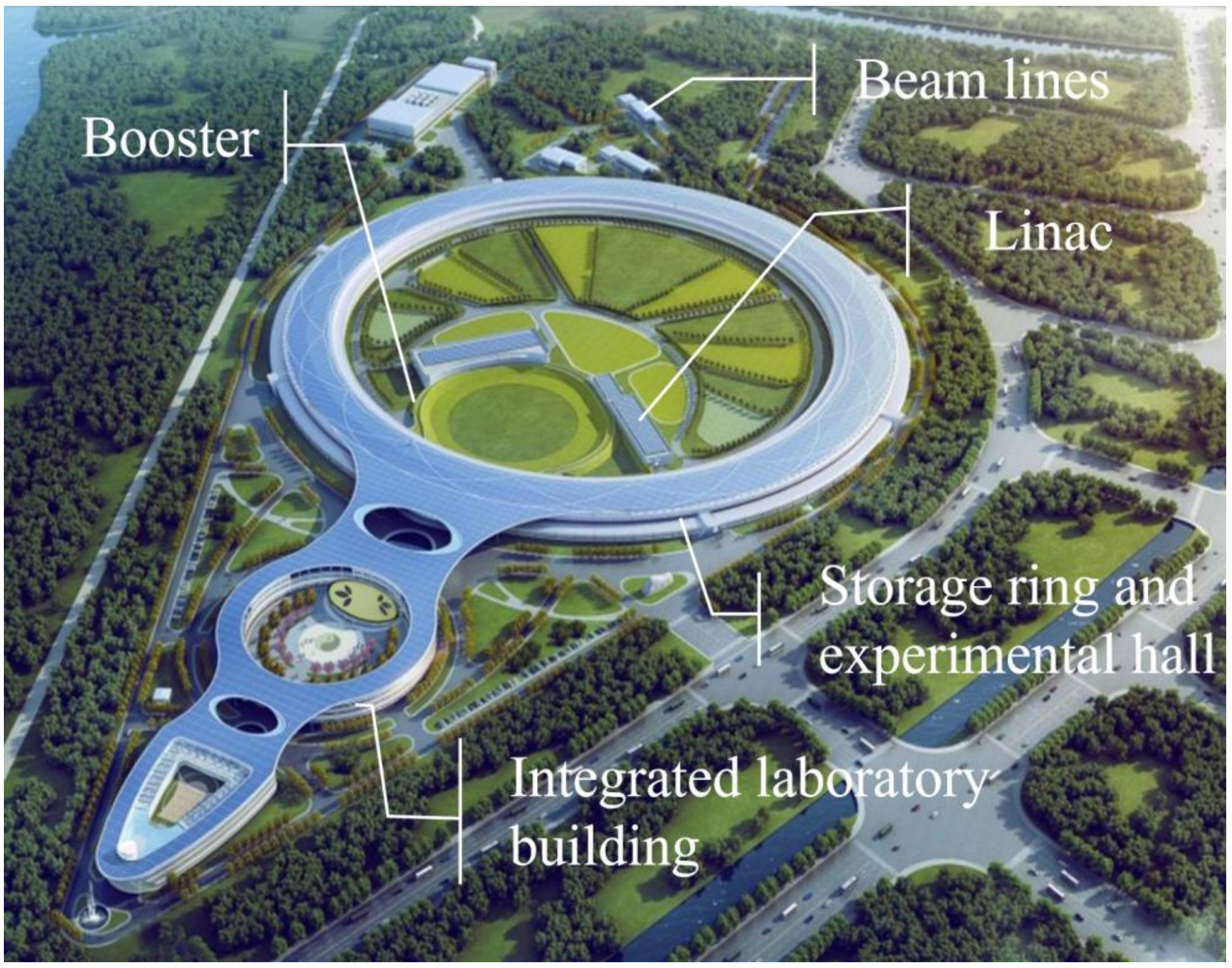

**Fig. 2** The layout of HEPS site

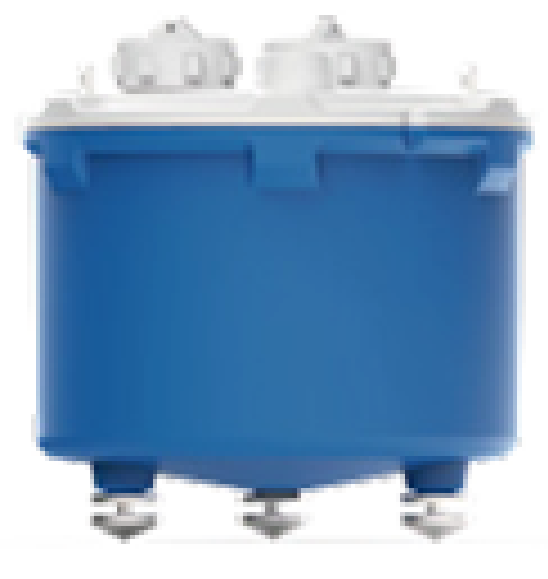

(a) Vibration test sensor

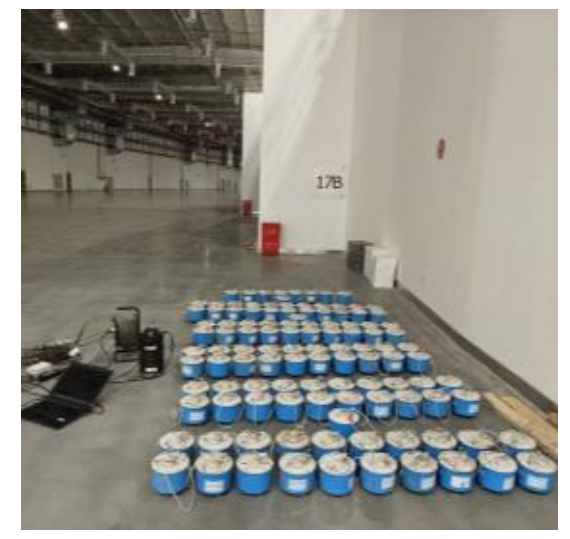

(b) On-site consistency test for the sensor

**Fig. 3** The sensor employed for testing and the on-site consistency test of the sensor on the HEPS experimental hall slab.

The vibration sensors used in this experiment were the Smartsolo IGU-16HR 3C AIO, a three-component velocimeter manufactured by Shenzhen Mianyuan Intelligent Technology Co., Ltd as shown in **Fig. 3** (a) [33]. These instruments have a frequency range of 0.2 Hz to 1000 Hz and a sensitivity of 76.7 V/m/s. Prior to testing, the consistency of all the sensors was verified on the slab of the HEPS experimental hall. The on-site test setup is depicted in **Fig. 2**(b). Random noise measurements were taken at midnight, and one hour of data was selected for consistent comparison, with a frequency resolution of 0.1 Hz. The displacement Root Mean Square (RMS) values, within the frequency range of 1-100 Hz, for all the sensors were calculated in the frequency domain using Equation (9) [34]:

$$x_{RMS}(f_1, f_2) = \sqrt{\sum_{f_1}^{f_2} S_x(f)\Delta f} = \sqrt{\sum_{f_1}^{f_2} \frac{S_v(f)}{(2\pi f)^2}\Delta f} \tag{9}$$

Where $S_x$ and $S_v$ are the displacement and velocity Power Spectrum Density (PSD) respectively, $f$ is the vibration frequency, $f_1$ and $f_2$ are the starting frequency and cutting frequency respectively considered, $\Delta f$ is the frequency resolution.

The comparative results of the displacement PSD for instruments S01 to S10 are illustrated in **Fig. 4** (a). All others sensors give similar results. The results demonstrate excellent consistency below 50 Hz across all sensors. The differences observed in the 50 Hz–100 Hz frequency range are primarily due to variations in the self-noise characteristics of the sensors. As shown in **Fig. 4**(b), the dispersion of the displacement RMS integral over the 1-100 Hz range for 64 units is within 0.5 nm. Furthermore, the comparison of the displacement RMS integral across the frequency range of 1-100 Hz for all 50 sensors also reveals excellent consistency, with a dispersion of within 0.5 nm.

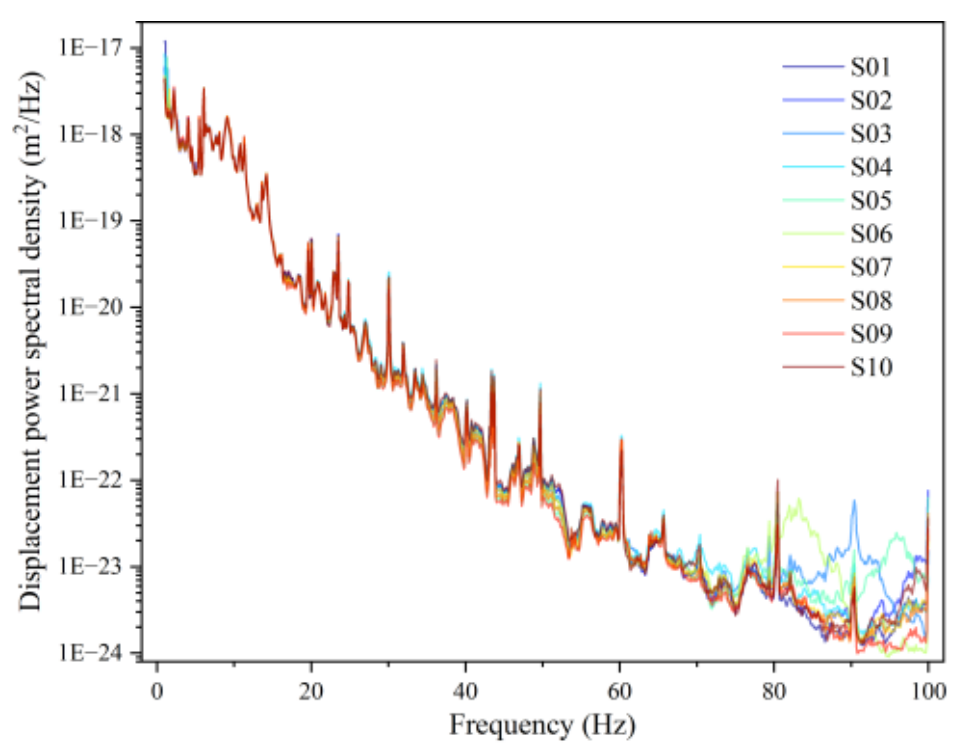

(a) Displacement PSD comparation of S01 to S10

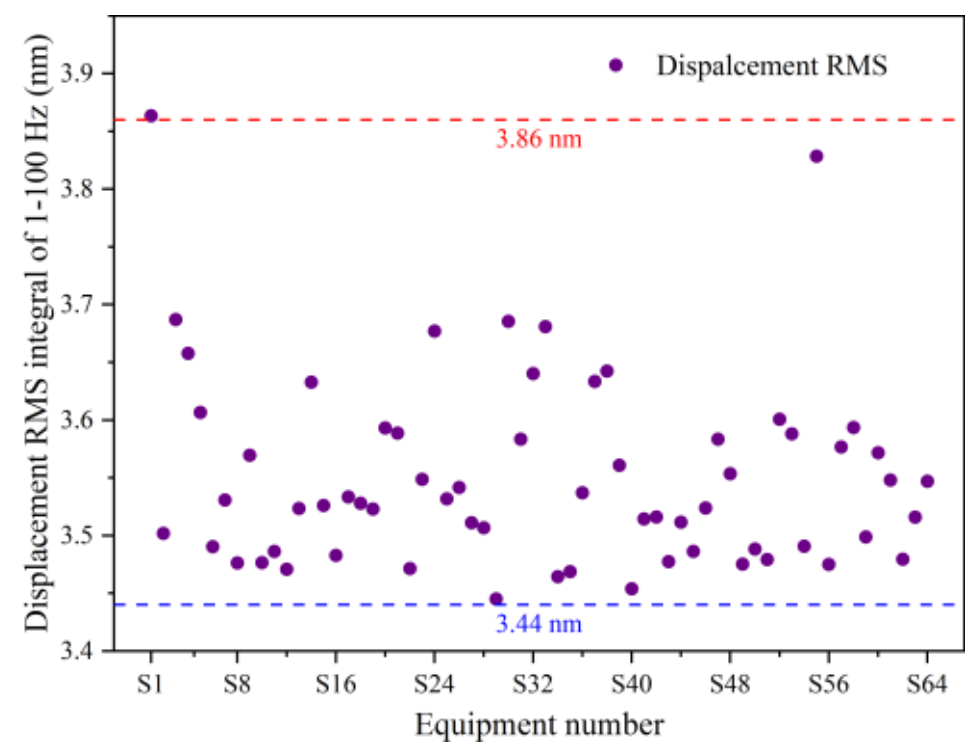

(b) The RMS distinguishes of all 64 sensors

**Fig. 4** Comparison of Displacement PSDs (S01–S10) and RMS Variability Across 64 Sensors

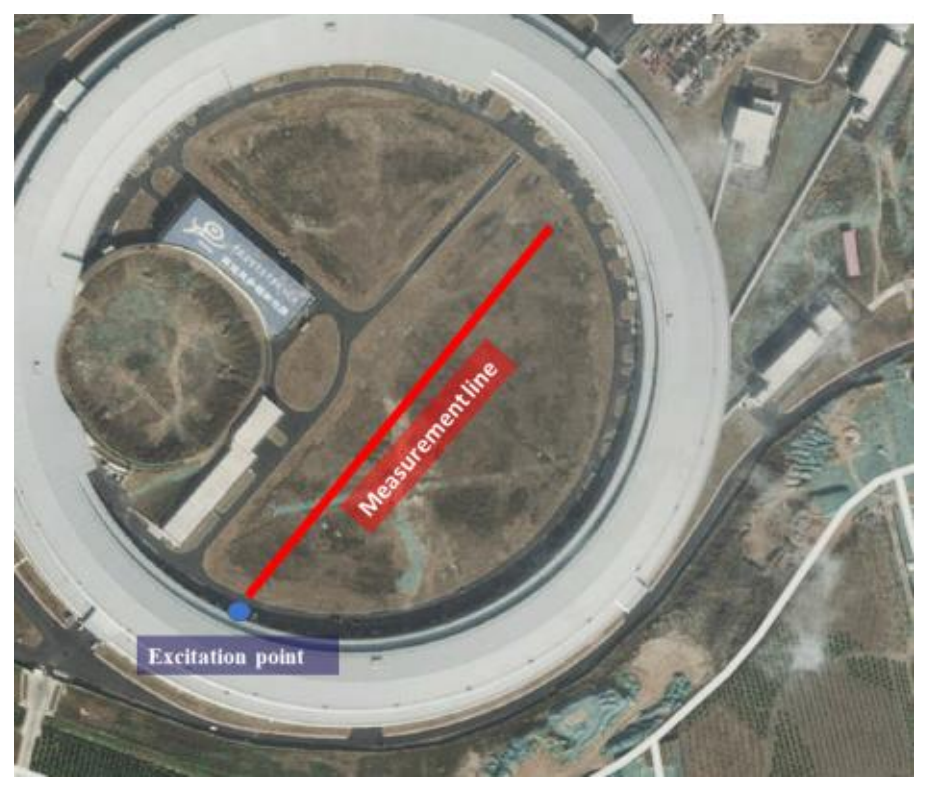

(a) The test location and the measurement line

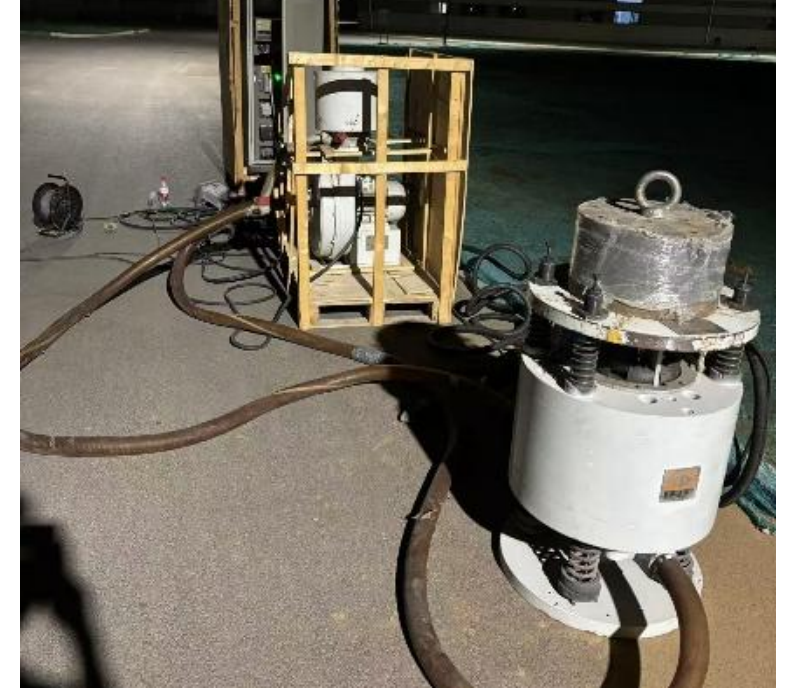

(b) Vibration exciter

**Fig. 5** The artificial vibration exciter and the excitation test location.

Vibration attenuation measurements were conducted on the bare ground within the HEPS storage ring, as shown in **Fig. 5**(a). The measurement line, marked in red, has a total length of 336 meters, with 64 measurement devices (labeled S1–S64) distributed along this line. The first 50 points are spaced approximately one meter apart to capture more detailed attenuation characteristics. Beyond this, the spacing gradually increases, ranging from about 10 meters to 100 meters, 20 meters to 200 meters, and 50 meters to 336 meters.

To generate unique vibration frequency, an artificial exciter was used. The shaker utilized in this study, shown in **Fig. 5** (b), is manufactured by Suzhou Dongling Vibration Testing Instrument Co. [35]. This shaker, with a total mass of 1 ton, generates wide-band vibrations from 1 Hz to 100 Hz. However, due to the limited maximum displacement of the shaker's moving coil during excitation, the vibration amplitude below 10 Hz cannot be set to a large amplitude. Despite this limitation, the signal-to-noise ratio remains favorable in the 10–100 Hz frequency range, even at a distance of 400 meters from the vibration source. This effectively mitigates the limitations associated with traditional artificial vibration sources, such as traffic loads and environmental noise.

The shaker generated vertical sinusoidal excitation forces ranging from 1 Hz to 100 Hz in 1 Hz increments, with each frequency lasting for 40 seconds. To mitigate interference from human activities, all tests were performed during midnight from 1am to 4am under low-ambient-noise conditions. During continuous excitation, the excitation frequency was altered every 40 seconds, with varying vibration amplitudes across frequencies. Transient amplitude fluctuations (1–2 seconds) occurred during frequency transitions. To minimize interference from these transients, the first 5 seconds and the last 5 seconds of data for each frequency interval were discarded in subsequent calculations. The intermediate 30-second stable segment was retained for analysis.

Although the vibration signal-to-noise ratio is generally high, certain measurement points at specific frequencies exhibit weak coherence with the initial testing point. In this study, the coherence between the signals from each sensor and the first sensor, ranging from S02 to S64, is calculated. Data with a coherence value below 0.9 are excluded. The calculation formula is as follows:

$$\gamma = \frac{\left|P_{xy}\right|^2}{P_{xx}P_{yy}} \tag{10}$$

Where $P_{xx}$ and $P_{yy}$ represent the auto power spectral densities of signal $x$ and signal $y$, respectively; $P_{xy}$

denotes the cross power spectral density between signal $x$ and signal $y$; and $\gamma$ represents the coherence. **Fig. 6** illustrates the coherence analysis for the last five measurement points of S59–S64. Data with coherence coefficients below the threshold of 0.9, observed at specific frequency intervals, were excluded in subsequent computational analysis to ensure fitting reliability.

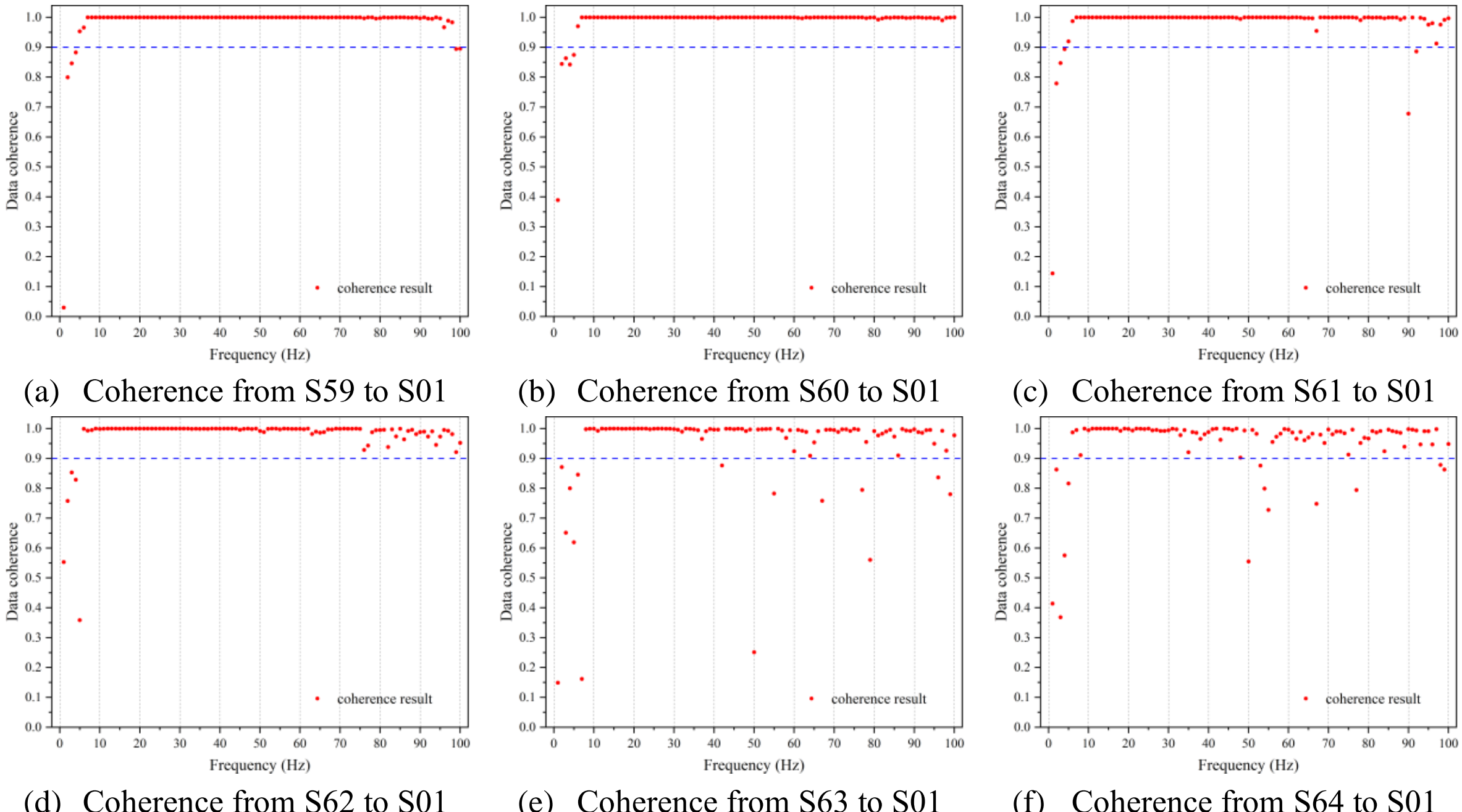

(a) Coherence from S59 to S01 (b) Coherence from S60 to S01 (c) Coherence from S61 to S01

(d) Coherence from S62 to S01 (e) Coherence from S63 to S01 (f) Coherence from S64 to S01

**Fig. 6** The coherence analysis for the last five measurement points of S59–S64

## 3.2. Formula discovery

Select the measuring point $r_0 = 2.1$m, which is the closest to the vibration source, as the reference point. With $n(f) = 0$ as the initial condition, the functional expressions of material damping coefficient $\alpha(f)$ and geometric damping coefficient $n(f)$ are iterated for 4 rounds according to the process shown in **Fig. 1**. The fitting results of each round are presented in **Table 1**.

**Table 1** The process of formula discovery

| Round | Material Damping Coefficient $\alpha(f)$ | Geometric Damping Coefficient $n(f)$ |
| --- | --- | --- |

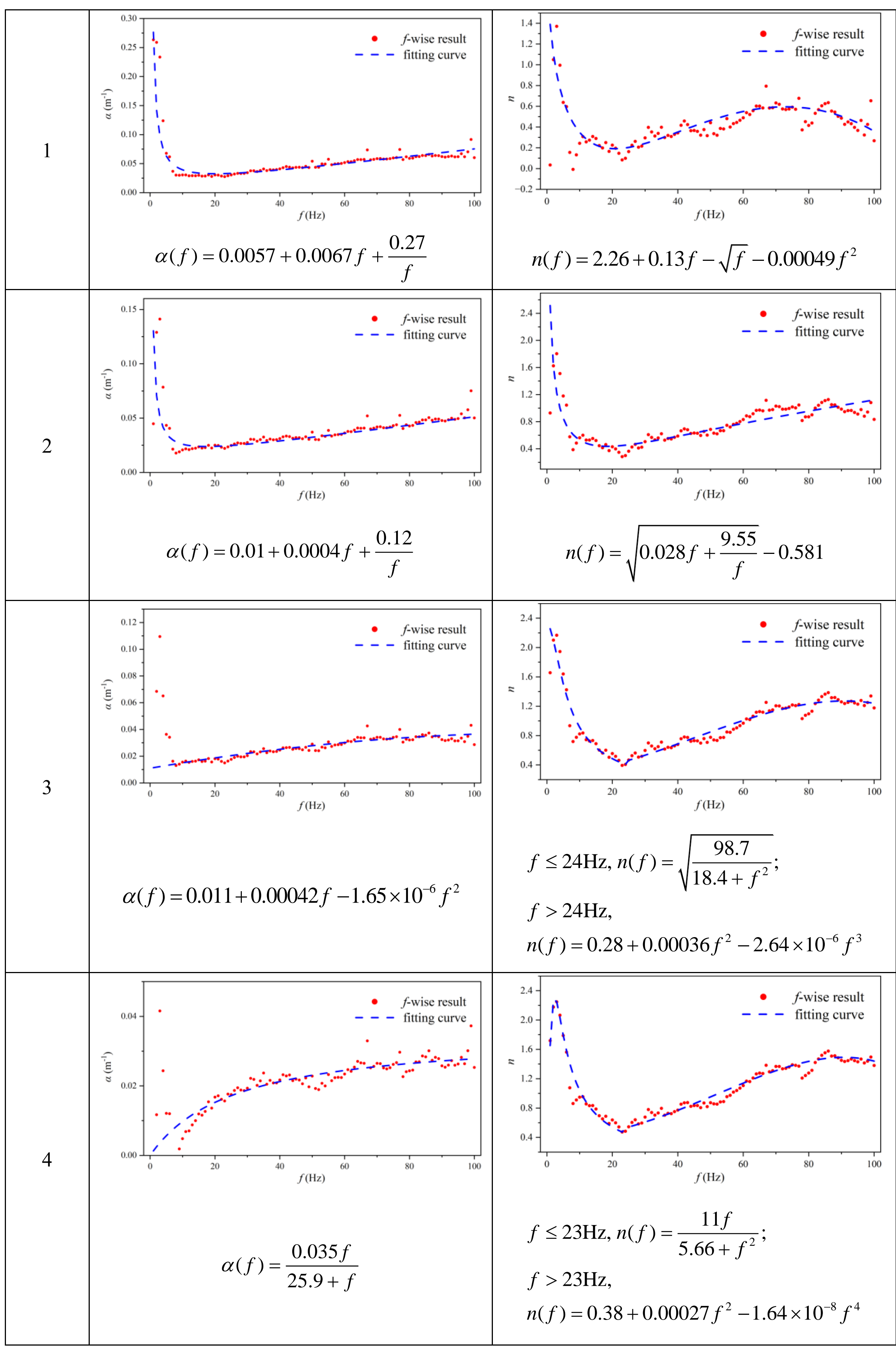

| | | |
|---|---|---|
| 1 | $\alpha(f)=0.0057+0.0067f+\dfrac{0.27}{f}$ | $n(f)=2.26+0.13f-\sqrt{f}-0.00049f^{2}$ |
| 2 | $\alpha(f)=0.01+0.0004f+\dfrac{0.12}{f}$ | $n(f)=\sqrt{0.028f+\dfrac{9.55}{f}}-0.581$ |
| 3 | $\alpha(f)=0.011+0.00042f-1.65\times10^{-6}f^{2}$ | $f\le 24\text{Hz},\ n(f)=\sqrt{\dfrac{98.7}{18.4+f^{2}}}$;<br>$f>24\text{Hz}$,<br>$n(f)=0.28+0.00036f^{2}-2.64\times10^{-6}f^{3}$ |
| 4 | $\alpha(f)=\dfrac{0.035f}{25.9+f}$ | $f\le 23\text{Hz},\ n(f)=\dfrac{11f}{5.66+f^{2}}$;<br>$f>23\text{Hz}$,<br>$n(f)=0.38+0.00027f^{2}-1.64\times10^{-8}f^{4}$ |

The fitting results of each iteration are substituted into Equation (5) to calculate the mean absolute error

(MAE) and relative mean absolute error (RMAE) of the formula, as shown in **Fig. 7**. Based on the formula discovered in the final iteration, the equation for determining the propagation and attenuation of ground vibration is as follows:

$$A_r = A_0\left(\frac{r_0}{r}\right)^{n(f)} e^{-\alpha(f)(r-r_0)} \tag{11}$$

$$n(f)=\begin{cases}\dfrac{11f}{5.66+f^2}, f\le 23\text{Hz}\\ 0.38+0.00027f^2-1.64\times10^{-8}f^4, f>23\text{Hz}\end{cases} \tag{12}$$

$$\alpha(f)=\frac{0.035f}{25.9+f} \tag{13}$$

where the MAE is $3.13\times10^{-7}$ $s^2$ and the RMAE is 90.47%. It can be observed that the iterative fitting process of the formula significantly reduces the fitting error. Moreover, as shown in **Table 1**, the intelligent formula generation model is capable of identifying formula forms that go beyond conventional empirical knowledge. This approach combines the powerful fitting ability of deep learning with the simplicity and intuitiveness of formulas, thereby providing a strong guarantee for the accuracy and interpretability of the model.

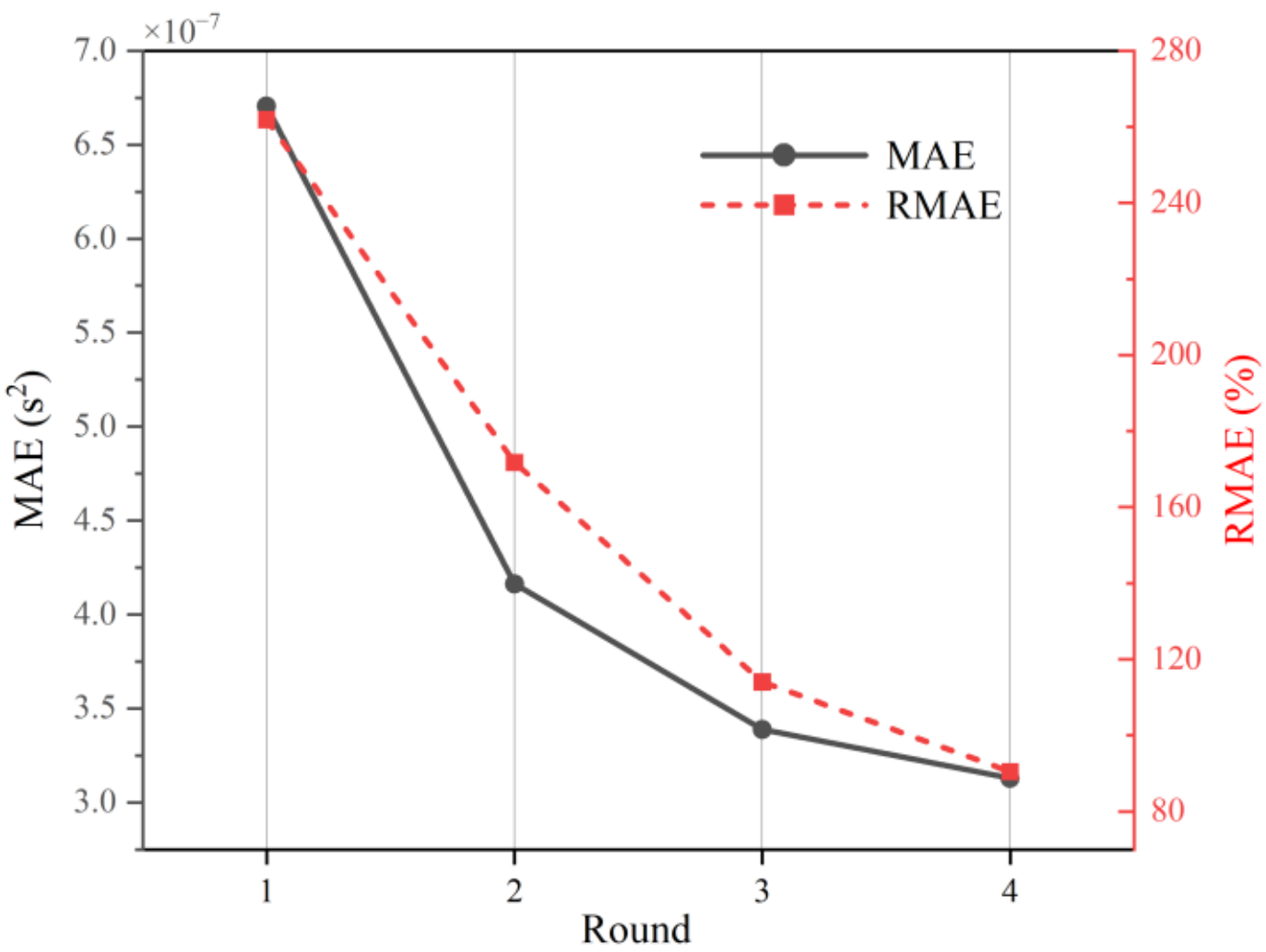

**Fig. 7** Error reduction during the iterative process

## 3.3. Physical rationality

In the Bornitz formula, the geometric damping coefficient $n$ is considered to be related to the spatial diffusion of wave energy, while the material damping coefficient $\alpha$ is associated with the absorption of wave energy by the soil. The material damping coefficient $\alpha$ is influenced by the complex properties of the soil, making it challenging to establish a universal physical interpretation. The study by Yang [36,37] suggests a positive correlation between the material damping coefficient $\alpha$ and frequency $f$, which aligns with the form of our equation. However, the relationship between the geometric damping coefficient $n$ and frequency $f$ has received relatively little attention in existing research.

The energy density of a wave is proportional to the square of its amplitude. Under ideal conditions, a point source in three-dimensional space emits waves that propagate spherically. According to the law of energy conservation, the total energy across different spherical surfaces must remain constant. Consequently, the energy density is inversely proportional to the square of the distance $r$ from the source, leading to the conclusion that the amplitude $A$ is inversely proportional to $r$:

$$A \propto r^{-1} \tag{14}$$

Similarly, for a point source emitting waves in a two-dimensional plane, the amplitude $A$ should be inversely proportional to $r^{0.5}$:

$$A \propto r^{-0.5} \tag{15}$$

Given the assumption that all energy on one plane is instantaneously transferred to the next, while in reality, only a portion of the energy is transmitted, the actual energy transfer is likely to be lower. Therefore, for ground vibrations that couple planar and spatial wave propagation, the geometric damping coefficient $n$ should be no less than 0.5. In our proposed Equation (12), the minimum value of the geometric damping coefficient $n$ is 0.47, which is consistent with the theoretical derivation mentioned above.

Without considering material damping, in an isotropic, homogeneous elastic medium, the displacement field $\mathbf{u}$ $(\mathbf{x}, t)$ satisfies the elastodynamic equation:

$$\rho \frac{\partial^2 \mathbf{u}}{\partial t^2} = (\lambda + 2\mu)\nabla(\nabla \cdot \mathbf{u}) - \mu \nabla \times (\nabla \times \mathbf{u}) \tag{16}$$

where $\rho$ is the mass density, $\lambda$ and $\mu$ are Lamé parameters characterizing the elastic properties of the medium. To investigate the relationship between geometric damping coefficient $n$ and $f$ or $r$, we employ the ABAQUS/Explicit solver, which is well-suited for transient dynamic analysis involving wave propagation. The explicit time integration scheme efficiently handles the high-frequency components of the elastic waves and ensures stable numerical simulations. As shown in **Fig.8**, we establish a 500m × 500m × 500m elastic solid domain to simulate the propagation of vibrations generated by a point source on the ground surface. The material properties are set as follows: density 1500 kg/m$^3$, Young's modulus 5000 kPa, and Poisson's ratio 0.4. A sinusoidal pulse with frequencies $f$ = 1 Hz, 5 Hz, 10 Hz, and 20 Hz is applied at the source location, and the maximum amplitude $A_\mathrm{m}$ at points in the horizontal observation direction is recorded. By analyzing the slope $k$ of the log( $A_\mathrm{m}$) - log( $r$) curve, the geometric damping coefficient $n$ for each frequency is determined, as illustrated in **Fig.9**.

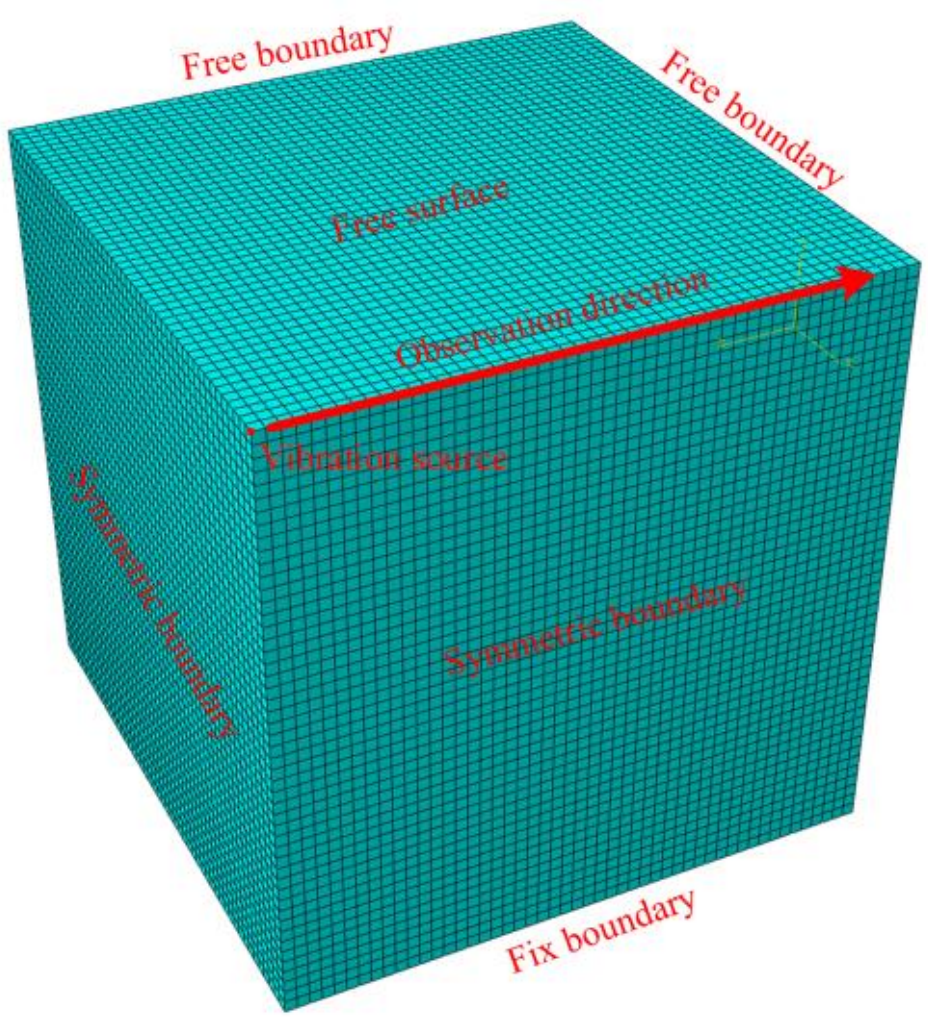

**Fig. 8** ABAQUS/Explicit modeling

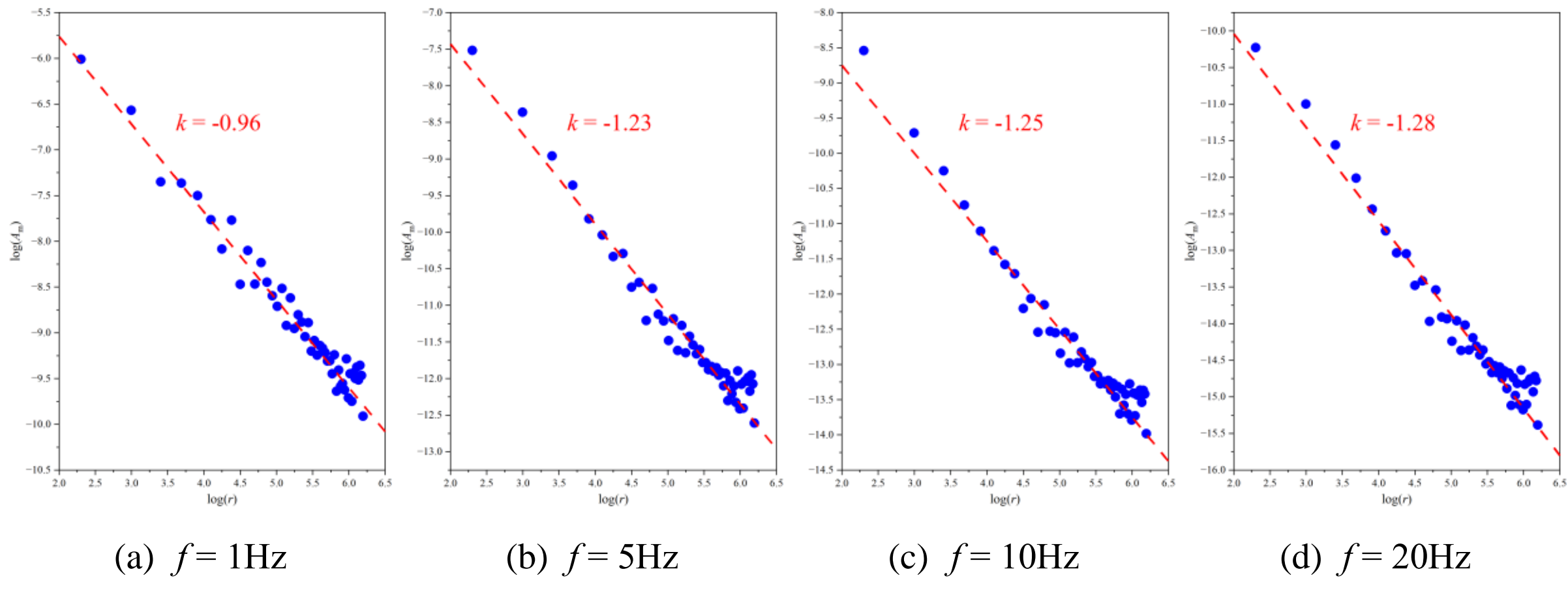

(a) $f$ = 1Hz (b) $f$ = 5Hz (c) $f$ = 10Hz (d) $f$ = 20Hz

**Fig. 9** Simulation results

The results indicate that the log($A_{\mathrm{m}}$) - log($r$) relationship is nearly a straight line for all tested frequencies, confirming that the geometric attenuation coefficient $n$ remains independent of distance $r$. This supports our assumption that the geometric damping coefficient $n$ depends only on frequency. Furthermore, the geometric damping coefficient exhibits an increasing trend with frequency, which is consistent with the second term in our proposed Equation (12). However, due to the complex and discrete nature of soil, accurately replicating its behavior in numerical simulations remains challenging. As a result, while our numerical simulations provide validation for the general form of our equation, the specific values of the geometric damping coefficients obtained from the simulations do not carry direct physical significance.

# 4. Analysis and comparison

## 4.1. Error analysis

Owing to the varying initial amplitudes of the vibration sources at different frequencies and the order-of-magnitude differences arising from the attenuation of vibrations with distance, the MAE, when measured in absolute terms, fails to adequately reflect the fitting error of the formula. Additionally, the complex nature of soil and the small measured true values at greater distances lead to an overestimation of the relative error. Consequently, the RMAE alone cannot intuitively represent the fitting performance of the formula. **Fig. 10** presents the cumulative distribution curve of the relative errors, where the x-axis represents the relative error, and the y-axis indicates the proportion of data points with a relative error less than or equal to the corresponding value out of the total number of data points. It can be observed that 36.5% of the data points have a relative error of less than 30%, 56.4% have a relative error of less than 50%, and 81.0% have a relative error of less than 100%. Additionally, 90.9% of the data points exhibit a relative error within 200%. Such precision is remarkably high for the fitting of soil problems, thereby demonstrating the effectiveness of our proposed method and formula. To further investigate whether the data points with relative errors exceeding 200% are due to the complex environmental conditions or to the inadequate fitting of the model at specific measurement points under certain frequencies, **Fig. 11** presents a distribution map of the data points with relative errors greater than 200%. It can be observed that the distribution of these data points does not exhibit any clear pattern (the concentration of these points in the lower range of $r$ is due to the fact that the measurement points are predominantly located in the lower range of $r$). This finding demonstrates the robust generalization capability of the formula within the tested range.

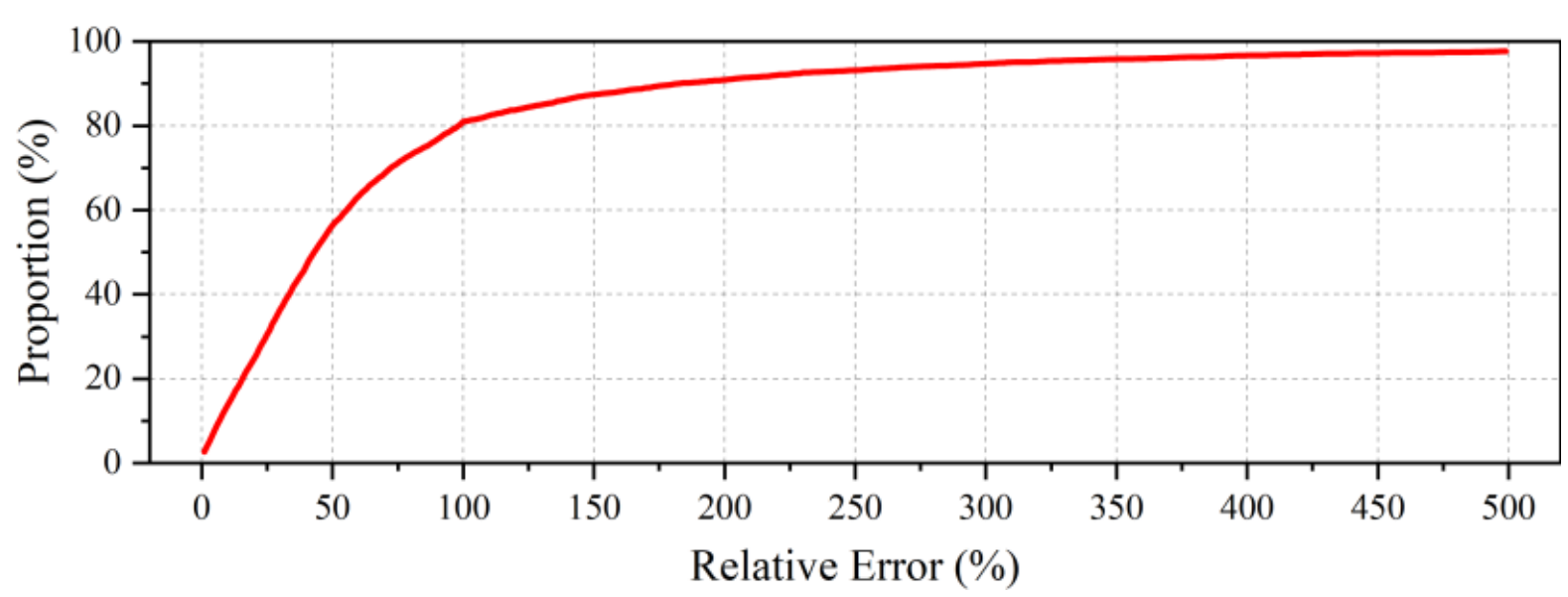

**Fig. 10** Cumulative distribution curve of the relative errors

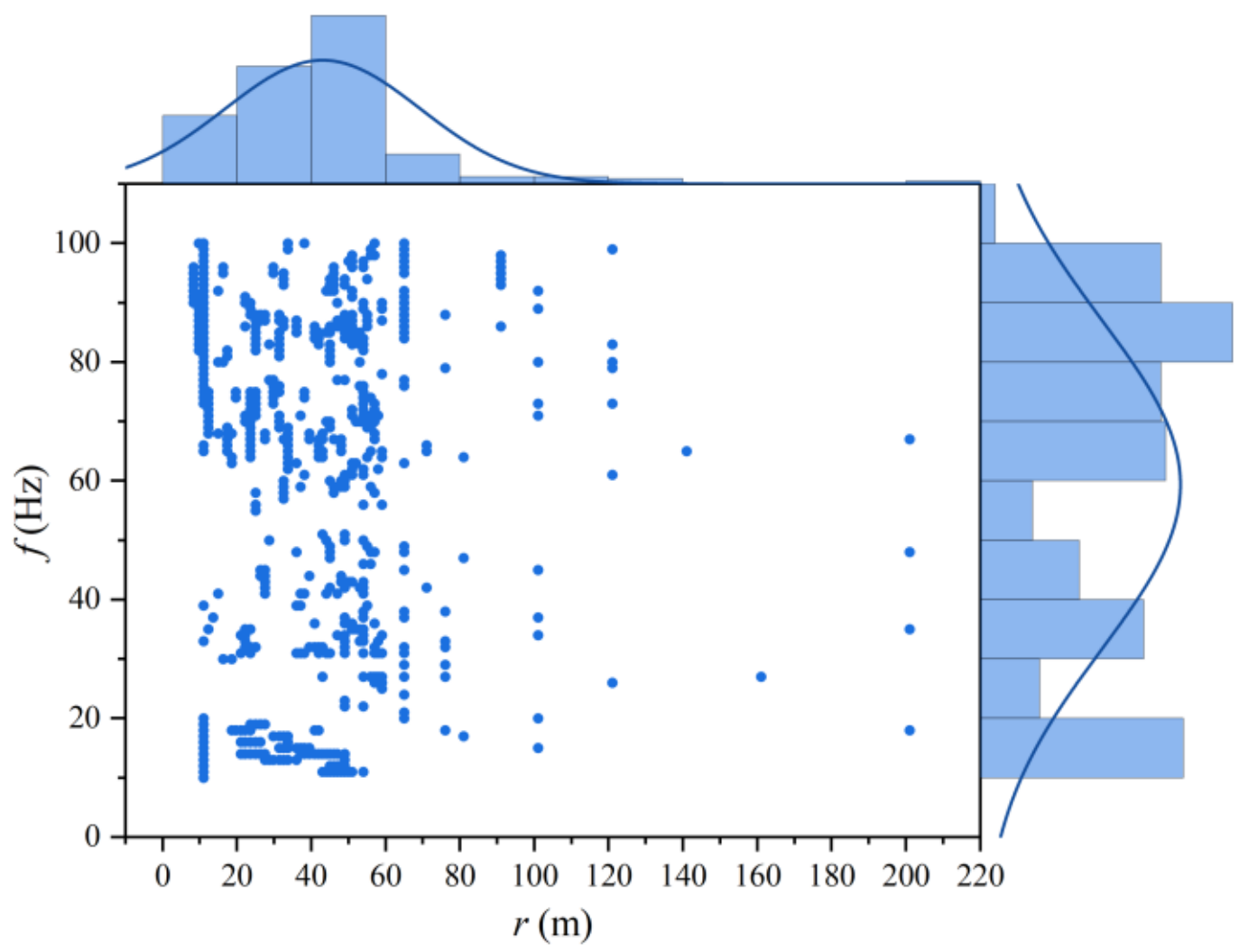

**Fig. 11** Distribution map of the data points with relative errors greater than 200%

Given the complex influence of environmental factors and the discrete nature of soil, the propagation and attenuation of ground vibration exhibit a certain degree of randomness. A single formula is insufficient to accurately assess their impacts. Assuming that the true amplitude $A_r$ at a distance $r$ from the vibration source and the predicted amplitude $\tilde{A}_r$ from the formula follow a $f$-independent probability distribution, the model can be calibrated using measurement results across different frequencies. The following three hypotheses were made regarding this probability distribution:

1. $A_r - \tilde{A}_r$ follows a normal distribution $\mathcal{N}(\mu,\sigma^2)$
2. $A_r / \tilde{A}_r$ follows a normal distribution $\mathcal{N}(\mu,\sigma^2)$
3. $A_r / \tilde{A}_r$ follows a log-normal distribution $\mathcal{LN}(\mu,\sigma^2)$, or equivalently, $\log(A_r / \tilde{A}_r)$ follows a normal distribution $\mathcal{N}(\mu,\sigma^2)$

The Shapiro-Wilk (SW) test is a widely used statistical method for assessing whether a dataset follows a normal distribution. It evaluates the null hypothesis ($H_0$) that the data originate from a normally distributed population. The test computes a test statistic $W$, which measures the extent to which the data conform to a normal distribution. The $W$ statistic is defined as:

$$W = \frac{\left(\sum_{i=1}^{n} a_i x_i\right)^2}{\sum_{i=1}^{n} (x_i - \overline{x})^2} \tag{17}$$

where $x_i$ represents the ordered sample values, $\overline{x}$ is the sample mean, and $a_i$ are constants derived from the expected values of order statistics from a standard normal distribution. The value of $W$ lies between 0 and 1, with values closer to 1 indicating a stronger conformity to normality. For comparative analysis, the $W$ statistic can be used to rank multiple datasets based on their degree of normality. A higher $W$ value suggests that the dataset more closely follows a normal distribution. Unlike the p-value, which only provides a binary decision regarding normality at a given significance level, the $W$ statistic allows for relative comparisons among different distributions. This makes it particularly useful in evaluating which dataset exhibits greater adherence to normality, especially in small-sample scenarios where other normality tests may be less reliable. Perform the SW test for each $r$ under the three aforementioned hypotheses and compute the corresponding $W$ values. The results are shown in **Fig. 12**.

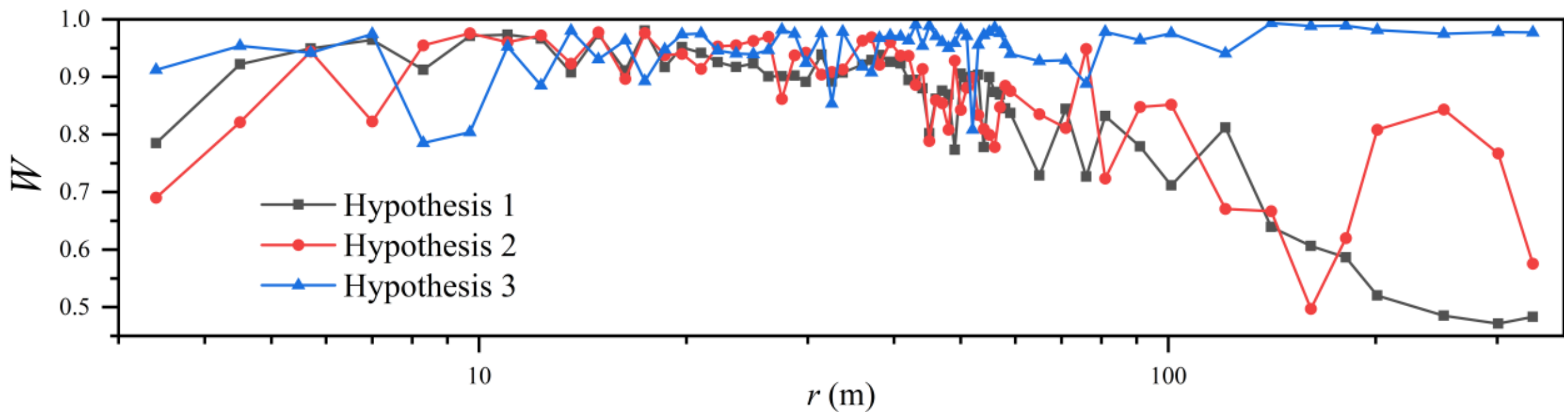

**Fig. 12** $W$ value for each $r$ under the three hypotheses

It can be observed that the overall $W$ values for Hypothesis 3 are higher, indicating that the assumption that $\log(A_r / \tilde{A}_r)$ follows a normal distribution is more reasonable. Building on this, for each $r$, we assume $\mu = 0$ and use data from different $f$ values to obtain an unbiased estimate of $\sigma$, as shown in **Fig. 13**. Subsequently, the intelligent formula generation model is employed to fit the functional relationship between $\sigma$ and $r$, expressed as follows:

$$\sigma = \frac{0.62}{1 - 0.0025r} \tag{18}$$

Based on the fitted results, the 1-2-3-$\sigma$ range for different $r$ values is determined, as illustrated in **Fig. 14**. The results show that 71.5% of the data points fall within the 1-$\sigma$ range, 94.2% of the data points fall within the 2-$\sigma$ range, and 98.7% of the data points fall within the 3-$\sigma$ range, demonstrating the effectiveness of the fitting model.

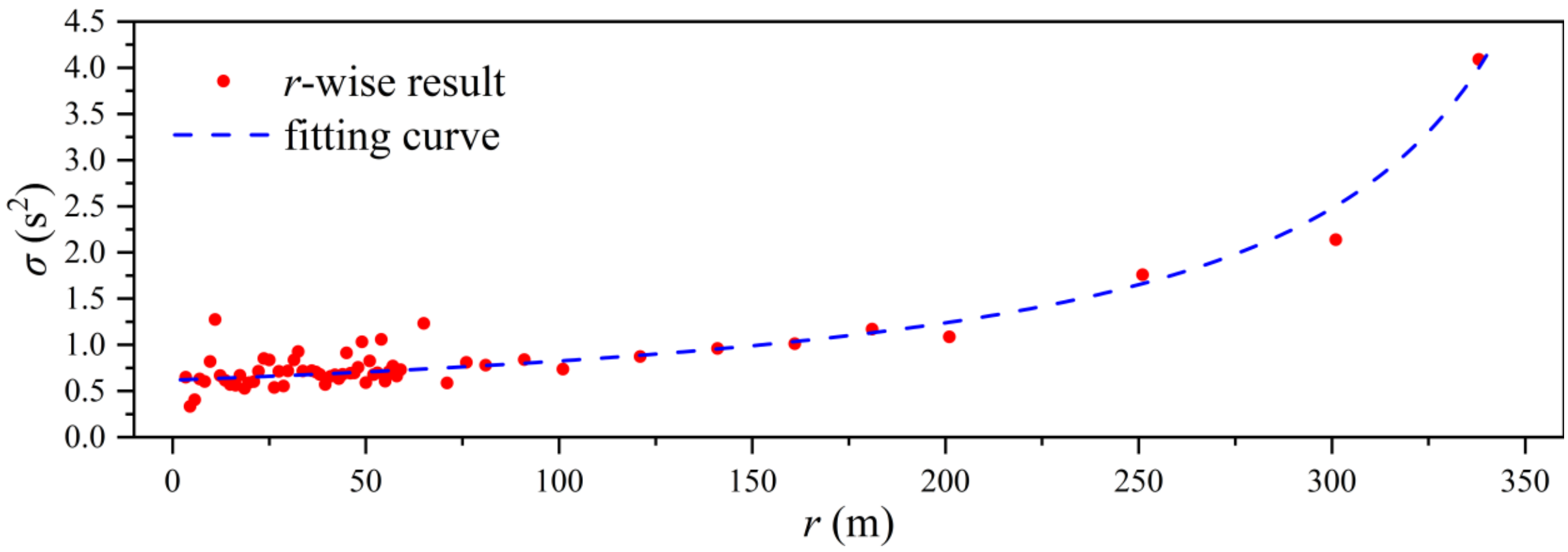

**Fig. 13** Estimate of $\sigma$ for each $r$

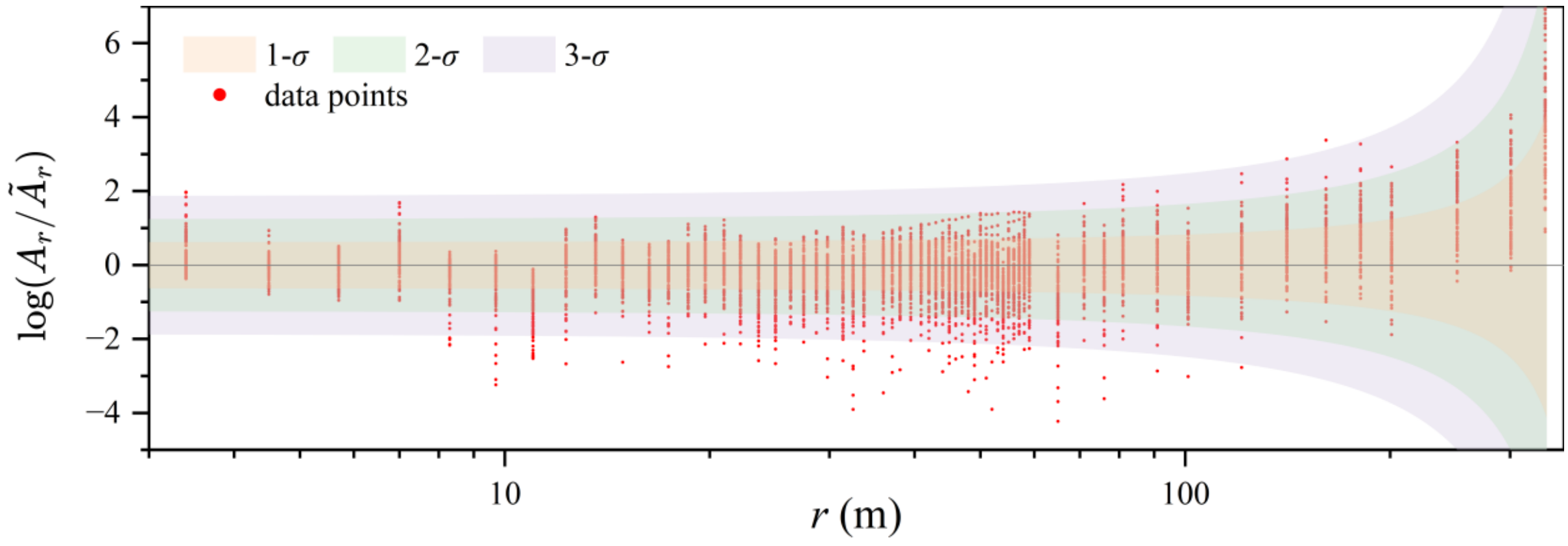

**Fig. 14** Error band for each $r$

## 4.2. Comparative evaluations

To verify the accuracy of the proposed formula in this study, we compare its performance with previously established formulas and the fitting results of black-box machine learning models. Regarding the previously proposed formulas, the first one is the original Bornitz formula, which does not consider the influence of

frequency:

$$A_r = A_0 \left( \frac{r_0}{r} \right)^n e^{-\alpha(r-r_0)} \tag{19}$$

This formula requires the calibration of the geometric damping coefficient $n$ and the material damping coefficient $\alpha$. The second is the frequency-dependent Bornitz formula developed by Yang [37] through theoretical derivation and experimental validation:

$$A_r = A_0 \sqrt{\frac{r_0}{r}[1-\xi(1-\frac{r_0}{r})]} e^{-\alpha f(r-r_0)} \tag{20}$$

which requires the calibration of the site coefficient $\xi$ and the material damping coefficient $\alpha$. The parameter calibration for both formulas is conducted using the *scipy.optimize.minimize* function [38] in Python library, aiming to minimize the mean absolute error (MAE). The final calibration results are as follows:

$$A_r = A_0 \left( \frac{r_0}{r} \right)^{0.885} e^{-0.071(r-r_0)} \tag{21}$$

$$A_r = A_0 \sqrt{\frac{r_0}{r}[1-0.865(1-\frac{r_0}{r})]} e^{-0.001f(r-r_0)} \tag{22}$$

Regarding black-box machine learning models, we selected two powerful fitting techniques: the traditional gradient boosting model Extreme Gradient Boosting (XGboost) and the recently emerging deep neural network (DNN). XGBoost is an efficient and scalable implementation of gradient boosting that has been widely used in regression and classification tasks. It builds an ensemble of decision trees sequentially, optimizing the model by minimizing a specified loss function while incorporating regularization to prevent overfitting. XGBoost is particularly well-suited for structured data and has demonstrated strong generalization capabilities across various domains. DNN is a class of artificial neural networks that consist of multiple hidden layers, enabling them to capture complex nonlinear relationships in data. Unlike traditional machine learning models, DNN leverages deep hierarchical feature extraction, making them highly effective for tasks requiring high-dimensional representations. In this study, we employ a feedforward DNN architecture trained using backpropagation and stochastic gradient descent. For the sake of fairness, similar to the formulas, both XGBoost and DNN take $A_0$, $r_0$, $r$, and $f$ as inputs, with $A_r$ as the output, employing MAE as the objective or loss function. Additionally, 75% of the data is used for training, and both the input and output variables are normalized using min-max scaling. In the experiment, the XGBoost is implemented with a maximum tree depth of 3, a learning rate of 0.1, and both the subsample ratio and column sampling ratio set to 0.8 to prevent overfitting. The model is trained for 1000 boosting rounds using the training dataset. The DNN consists of four hidden layers, each with 32 neurons, utilizing the sigmoid activation function. The model is optimized using the Adam optimizer with an initial learning rate of 0.01 and is trained for 10000 epochs.

To compare the accuracy of different methods in fitting the data, the results are presented in **Table 2**. It can be observed that the proposed formula achieves the lowest MAE and RMAE among all formula-based models. In contrast, the black-box machine learning models, XGBoost and DNN, exhibit lower MAE values, which is consistent with their strong fitting capabilities. However, both models have significantly higher RMAE values compared to the formula-based approach, indicating that their predictions are more susceptible to the influence of extreme values. However, as analyzed in **Section 4.1**, MAE and RMAE alone are insufficient to comprehensively evaluate the effectiveness of data fitting. To provide further insight, **Fig. 15** illustrates the cumulative distribution curves of relative errors for different methods. It can be observed that the proposed formula exhibits a significant advantage over previous formulas within the range of relative errors below 100%, achieving a fitting accuracy comparable to that of black-box models. Additionally, the cumulative distribution curves of Bornitz's and Yang's formulas show a sudden increase at the 100% mark. This indicates that their predicted values are often significantly lower than the actual values, leading to a large number of points with relative errors close to 100%. This phenomenon is further confirmed in **Fig. 16**, which presents the distribution of the ratio between predicted $\tilde{A}_r$ and actual values $A_r$ for different methods. **Fig. 16** also reveals that some of the predicted values from XGBoost and DNN are negative, which is physically unrealistic. This further highlights the issue of low interpretability in black-box models, which becomes particularly pronounced in physics-related problems. Considering various evaluation criteria, our proposed method effectively integrates the strong fitting capability of black-box models with the simplicity and intuitiveness of analytical formulas. The discovered formula strikes a balance between accuracy and interpretability, demonstrating significant practical value for ground vibration prediction.

**Table 2** Comparison of fitting accuracy of different methods

| Method | MAE ($\times 10^{-7}s^2$) | RMAE (%) |
|---|---|---|
| Proposed | 3.13 | 90.47 |
| Bornitz's | 4.16 | 99.26 |
| Yang's | 3.69 | 186.08 |
| XGboost | 1.25 | 567.90 |
| DNN | 1.74 | 701.52 |

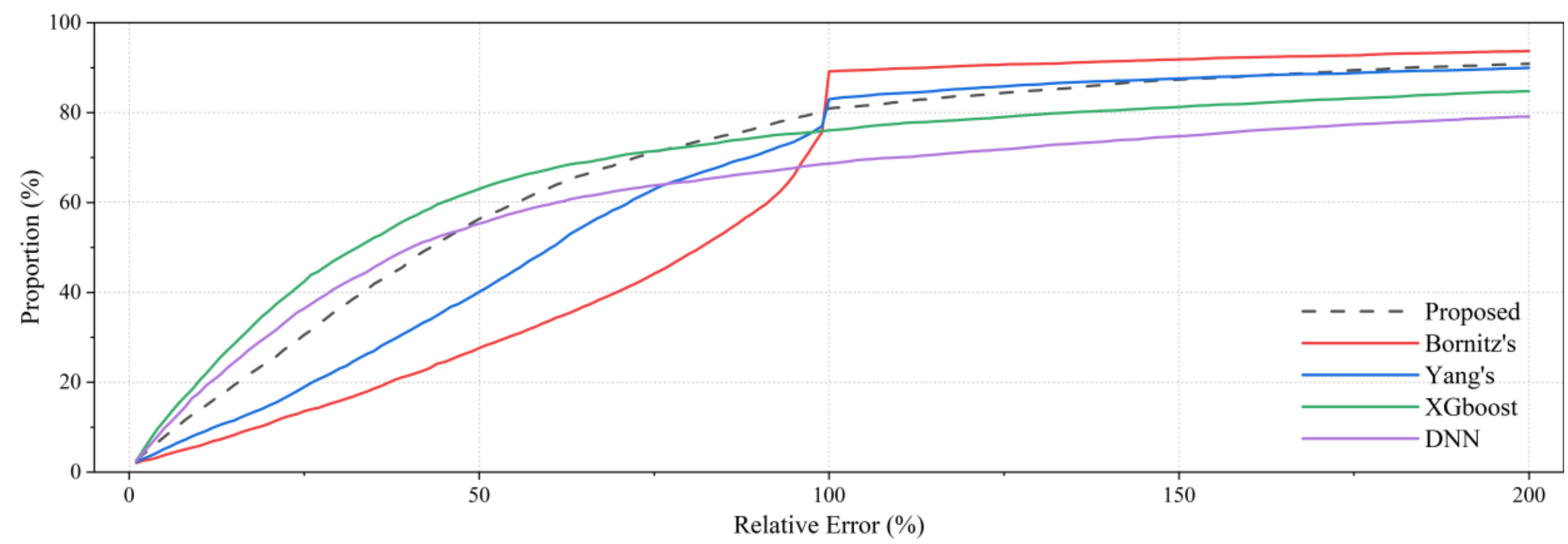

**Fig. 15** Cumulative distribution curve of the relative errors for different methods

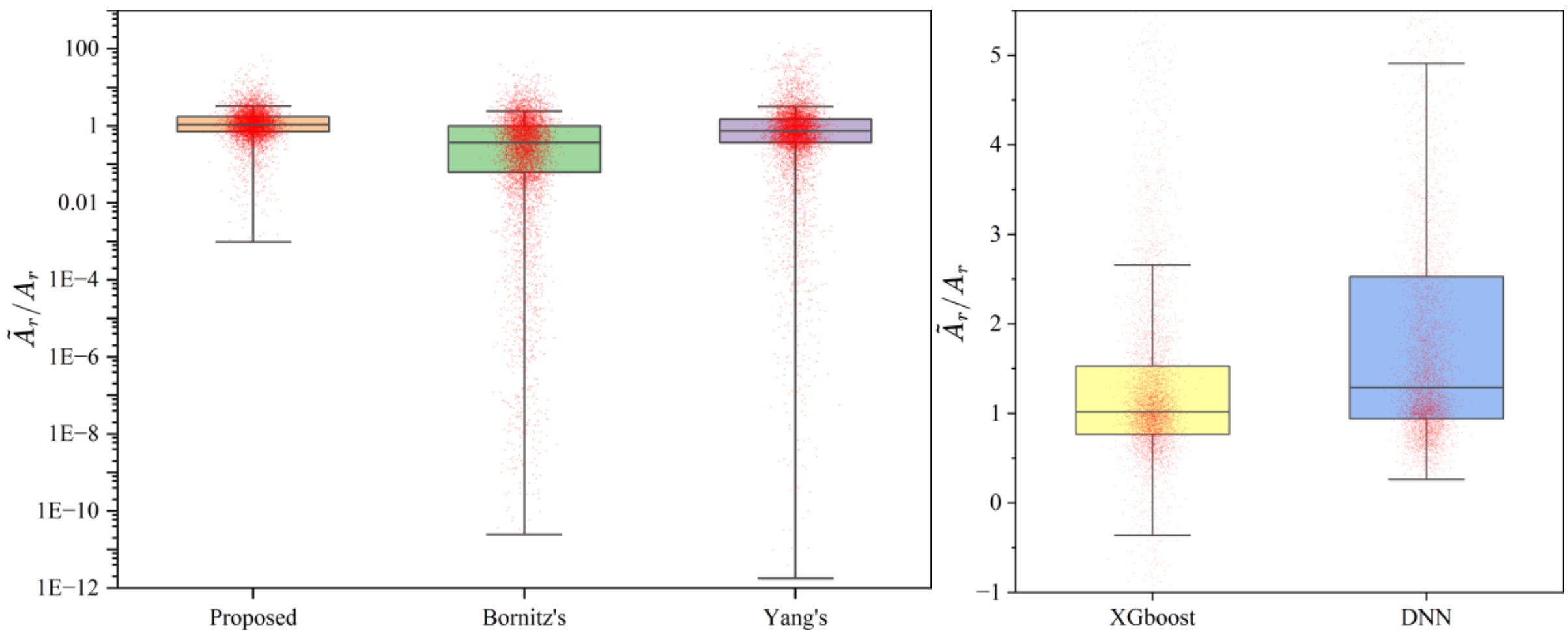

**Fig. 16** Distribution of the ratio between predicted and actual values for different methods

# 5. Conclusions

In this study, we propose an intelligent method for fitting analytical formulas that describe ground vibration propagation and attenuation. The method is validated through surface vibration measurements conducted at the HEPS in Beijing. The main contributions of this work are summarized as follows:

1. A formula discovery method integrating traditional machine learning techniques with intelligent formula generation models was proposed. This approach enables the efficient and rapid derivation of propagation and attenuation formulas based on the Bornitz model.
2. High-fidelity vibration experiments are presented at the HEPS site, providing reliable data for understanding ground vibration propagation. The results lay a solid foundation for evaluating vibration impacts on precision instruments.
3. Using the proposed method, a propagation attenuation formula suitable for ground vibrations at the HEPS site was discovered. The validity of the formula's structure was justified through theoretical analysis and numerical modeling.
4. A probabilistic model was employed to analyze the errors of the discovered formulas, providing a bounded error range model.

5. A comparative analysis was performed against previously proposed formulas and black-box models, demonstrating that the proposed method achieves both high accuracy and interpretability, making it suitable for ground vibration prediction and assessment.

This research offers a practical and scalable methodology for quantitatively evaluating the impact of environmental vibrations—such as those induced by traffic, construction, or natural events—on high-precision scientific infrastructure. The derived propagation formulas, with both physical interpretability and predictive accuracy, hold great promise for supporting the design, operation, and vibration control strategies of next-generation ultra-sensitive facilities, including synchrotron light sources, quantum laboratories, and gravitational wave observatories.

## 6. Acknowledgements

The authors gratefully acknowledge the financial support provided by the General Program of National Natural Science Foundation of China, under fund approval No. of 12375150.